\documentclass[11pt, a4paper]{article}

\usepackage{a4wide}
\usepackage{amsmath}
\usepackage{amsfonts}
\usepackage{enumerate}
\begin{document}
\title{Multi-Use Multi-Secret Sharing Scheme for General Access Structure
}
\author{
Partha Sarathi Roy and Avishek Adhikari \\
Department of Pure Mathematics\\
University of Calcutta\\
35, Ballygunge Circular Road, Kolkata 700019, India\\
E-mail : royparthasarathi0@gmail.com \& aamath@caluniv.ac.in
}
\date{}	
\maketitle

\newcommand{\pf}{{\bf Proof : }}
\begin{abstract} The main aim of this paper is to construct a multi-secret sharing scheme for general access structure in a trusted dealer model using suitable hash function and Lagrange's interpolation method. Even though, the proposed scheme is a multi-secret and multi-use one, each participant has to carry only one share. The suitable use of collision resistant one way hash function makes the scheme efficient and multi-use. Moreover, the scheme has a nice property that secrets, participants or qualified sets of participants may be added to or even may be made inactive dynamically by the dealer to get a new access structure without altering the shares of the existing participants in the old access structure. Finally, in the proposed scheme, both the combiner and the share holders can verify the correctness of the information that they are receiving from each other. 

\end{abstract} 



\section{Introduction} 
In the open system environment, it is important to restrict access of confidential information in the system or on certain nodes in the system. Access is gained through a key, password or token and governed by a secure key management scheme. If the key or the password is shared among several participants in such a way that it can be reconstructed only by a significantly large and responsible group acting in agreement, then a high degree of security is attained.
Also in many information technology applications, as well as in real world, it is desirable that actions or secrets to be protected by more than one key (jointly or separately) or that there be several keys and more than one way to recover the secret to initiate the action, using different combinations of keys.

Shamir \cite{sha} and Blakley \cite{bla} independently addressed this problem in 1979 when they introduced the concept of a threshold scheme. A $(t, n)$ {\em threshold scheme} is a method where $n$ pieces of information of the secret key $K$, called {\em shares} are distributed to $n$ participants so that the secret key can be reconstructed from the knowledge of any $t$ or more shares and the secret key
can not be reconstructed from the knowledge of fewer than $t$ shares.

But in reality, there are many situations in which it is desirable to have a more flexible arrangement for reconstructing the secret key. Given some $n$ participants, one may want to designate certain authorized groups of participants who can use their shares to recover the key. This kind of scheme is called secret sharing scheme for general access structure \cite{ben}, \cite{ito}.

\noindent Formally, a {\em secret sharing scheme for general access structure} is a method of sharing a secret $K$ among a finite set of participants $\mathcal{P}= \{P_1,P_2, \ldots, P_n \}$ in such a way that
\begin{enumerate}
\item if the participants in $\mathcal{A} \subseteq \mathcal{P}$ are qualified to know the secret, then by pooling together their partial information, they can reconstruct the secret $K$,
\item  any set $\mathcal{B} \subset \mathcal{P}$ which is not qualified to know $K$, cannot reconstruct the secret $K$.
\end{enumerate}

The key is chosen by a special participant $\mathcal{D}$, called the {\em dealer}, and it is usually assumed that $\mathcal{D} \notin \mathcal{P}$. The dealer gives partial information, called {\em share}, to each participant to share the secret key $K$. In some schemes, there is another special participant, called the {\em combiner}, to whom the participants give their shares to get the corresponding secret.
The collection of subsets of participants that can reconstruct the secret in this way is called {\em access structure} $\Gamma$.

Initially, in all the secret sharing schemes, it  was assumed that the dealer, the participants and the combiner are all trusted. Under this assumption, many secret sharing schemes for threshold as well as for general access structures were proposed \cite{sha}, \cite{aa}. These concepts were generalized by secret sharing schemes for general access structure with more than one secret with an aim to use the same share of a particular participant more than once. Those schemes are known as {\em multi-secret sharing schemes} for general access structure.  But, it was pointed out that if the combiner is not trusted, the above schemes may not be safe to use more than once under the following situation. Suppose a particular participant, say $P$, holding only one share, is a member of two different qualified sets of participants having two different secrets. Now to reveal the first secret, the participant $P$ has to give the share to the combiner. Having the share of $P$, the combiner may play the role of $P$, without the knowledge of $P$, while reconstructing the 2nd secret corresponding to the 2nd qualified set of participants. To overcome this problem, in 1994, He-Dawson \cite{he} claimed to propose a multi-stage threshold secret sharing scheme to share multiple secrets based on collision resistant one-way hash function. But in 2007, Geng $et$ $al.$ \cite{geng} pointed out that He-Dawson scheme was actually the one-time-use scheme. In \cite{geng}, they proposed a new threshold multi-use multi-secret sharing scheme. In 2006, Pang $et$ $al.$ \cite{pang} proposed a multi-secret sharing scheme, based on two variable one-way function and Lagrange's interpolation method, for general access structure in which all the secrets are revealed at a time. In 2008, Wei $et$ $al.$ \cite{wei} proposed a multi-stage secret sharing scheme, based on Lagrange's interpolation method and intractability of discrete logarithm problem, for general access structure in which secrets reveal in a predetermined order.  But, in most of the cases, it is required to share different secrets with different access structures. 

For more practical purpose, a secret sharing scheme should have the property that the access structure may be modified dynamically i.e., the secrets, participants or qualified sets of participants may be added to or may be made inactive in the existing access structure to get a new access structure without altering the shares of the existing participants in the old access structure. This type of scheme is known as {\em renewable} secret sharing scheme.

So, in the current scenario, a multi-use, multi-secret, renewable, verifiable secret sharing scheme for general access structure is essential. In the current paper, we deal with all the above aspects of the secret sharing scheme.

The rest of this paper is organized as follows. In Section \ref{1.2},  a new scheme having all the properties mentioned above is introduced. In Section \ref{1.3},  renew process of the proposed scheme is studied. The analysis of the proposed scheme is given in Section \ref{1.4}. Finally, Section \ref{1.5} deals with the conclusion of the paper.

\section{A New Multi-Secret Sharing Scheme for General Access Structure}
\label{1.2}
In this section, a new renewable, multi-use, multi-secret sharing scheme for general access structure in a trusted dealer model is introduced using suitable hash function \cite{stin} and Lagrange's interpolation method.

\noindent{\bf Aim of the scheme} \\
Let the dealer want to share $k$ secret integers $s_1$, $s_2$, $\ldots$, $s_k$ among $n$ participants $P_1$, $P_2$, $\ldots$, $P_n$ in such a way that corresponding to each secret $s_i$, there exists an access structure $\Gamma_{s_{i}}=\{{\mathcal{A}_{1}^{s_{i}}}, {\mathcal{A}_{2}^{s_{i}}}, \ldots, {\mathcal{A}_{l_{i}}^{s_{i}}}\}$, where ${\mathcal{A}_{q}^{s_{i}}}=\{P_{1}^{i_q}, P_{2}^{i_q}, \ldots, P_{m_{i_{q}}}^{i_q}\}$$\subseteq$$\{P_{1}$, $P_{2}$, \ldots, $P_{n}\}$, $|{A_{q}^{s_{i}}}|\geq 2$, $q=1, 2, \ldots,l_i,$  $i=1, 2, \ldots, k$ and $1\leq l_{i}\leq 2^{n}-(n+1)$. Note that we assume all the $s_i$'s to be non negative integers as otherwise the dealer may add some suitable positive integer to all the $s_i$'s to transfer them into non negative integers and may publish the value of the fixed positive integer in the public domain. Also, note that for some $c, d\in \{1, 2, \ldots, k\}$ with $c\neq d$ it may happen that $\Gamma_{s_{c}} \cap \Gamma_{s_d}\neq \phi$. To obtain such scheme, we explain the following four different phases namely, the Dealer Phase, the Participant Phase (I), the Combiner Phase and the Participant Phase (II).   

\subsection{Dealer phase}
{\bf(I) Initialization stage}
\begin{enumerate}
\item The dealer chooses the following.  
\begin{itemize}
\item $p$, a prime such that $s_i <p$ and $n<p$, $i=1,2,\ldots, k$;
\item $h$, a secure collision resistant one-way hash function which takes as input a binary string of any length and provides as output a binary string of fixed length $[\log_2 p]+1$. 
\item $ID_{j}$, the distinct identifier corresponding to each of the participant $P_{j}$, $j=1, 2, \ldots, n$, where $ID_{j} \in_R \mathbb {Z}_{p}^*=\mathbb{Z}_{p} \setminus \{ 0\}$, where $``\in_R"$ denotes the random selection.
\end{itemize} 
\item All the above listed entities and the access structure are made public by the dealer.
\end{enumerate}

\noindent {\bf (II) Pseudo share generation stage} \\ \\
To generate the pseudo shares and to distribute the shares to each participant the dealer performs the following:
\begin{enumerate}
\item The dealer chooses distinct $x_{j} \in_{R}\mathbb{ Z}_{p}$ and sends it secretly to each of the  participants $P_{j}$, where $j=1,2, \ldots, n$. This step may also be performed in another way. Each participant can choose his or her share by himself or herself and can send it to the dealer through a secure channel. The dealer keeps on asking the shares from the participants, till all the shares are distinct. 

\item For the $q$th qualified set of $\Gamma_{s_i}$, the dealer chooses $d_{1}^{i_q}, d_{2}^{i_q}, \ldots, d_{m_{i_{q}}-1}^{i_q}\in_R \mathbb{ Z}_{p}$ to construct the polynomial 
$f_{q}^{s_{i}}(x)=s_{i}+d_{1}^{i_q}x+d_{2}^{i_q}x^{2}+\ldots +d_{m_{i_{q}}-1}^{i_q}x^{m_{i_q}-1}$,  $i=1, 2, \ldots, k,$ $ q=1, 2,\ldots, l_{i}.$

\item Let $l= max \{l_1, l_2, \ldots, l_k\}, u= [\log_{2}k]+1,  v= [\log_{2}l] +1$. For the participant $P_{b}^{i_q}\in \mathcal{A}_{q}^{s_{i}}$ in $\Gamma_{s_{i}}$, the dealer computes the {\em pseudo share} $\mathcal{U}_{P_{b}^{i_q}}=h(x_{P_{b}^{i_q}}||i_u||q_v)$, where $i=1, 2, \ldots, k, $ $ q=1, 2, \ldots, l_{i}, $ $ b=1, 2, \ldots, m_{i_{q}}$, $i_u$ is the $u$-bit binary representation of $i$ and $q_v$ is the $v$-bit binary representation of $q$. Here, $``||"$ denotes the concatenation of two binary strings. Note that, though $x_{P_{b}^{i_q}}$ is an element of $\mathbb{Z}_p$, to avoid the notational complexity, we use the same notation to represent the binary representation of $x_{P_{b}^{i_q}}$. Applying the similar argument, we consider $\mathcal{U}_{P_{b}^{i_q}}$ as an element of $\mathbb{Z}_p$.

\item The dealer computes  $\mathcal{B}_{P_{b}^{i_q}}=f_{q}^{s_{i}}(ID_{b}^{i_q})$, where $i=1, 2, \ldots, k,$  $ q=1, 2, \ldots, l_{i},$  $ b=1, 2, \ldots, m_{i_{q}}.$

\item Finally, the dealer computes and publishes $\mathcal{M}_{P_{b}^{i_q}}=(\mathcal{B}_{P_{b}^{i_q}}-\mathcal{U}_{P_{b}^{i_q}})$, where $i=1, 2, \ldots, k, $ $ q=1, 2, \ldots, l_{i}$,  $ b=1, 2, \ldots, m_{i_{q}}.$

\end{enumerate}

\noindent{\bf (III) Prerequisites for verification stage} \\ \\
In this stage, the dealer prepares all the prerequisites for the verification of the participants by the combiner and  the verification of the combiner by the participants.
\begin{enumerate}
\item For the verification of the participants by the combiner, the dealer computes and publishes   $\mathcal{N}_{P_{b}^{i_q}}=h({\mathcal{U}_{P_{b}^{i_q}}})$, where $i=1, 2,\ldots, k$, $ q=1, 2,\ldots , l_{i}$,  $b=1, 2,\ldots, m_{i_{q}}.$ 
\item For the verification of the combiner by the participants, the dealer computes and publishes $S_{i}=h({s_{i}})$, where $i=1, 2,\ldots,k.$
\end{enumerate}

\subsection{Participant Phase (I)} 
Let all the members of ${\mathcal{A}_{q}^{s_{i}}}=\{P_{1}^{i_q}, P_{2}^{i_q}, \ldots, P_{m_{i_{q}}}^{i_q}\}$ in $\Gamma_{s_i}$ accumulate to reveal $s_i$. Each participant $P_{b}^{i_q}$ of $\mathcal{A}_{q}^{s_{i}}$ in $\Gamma_{s_i}$, $b=1, 2, \ldots, m_{i_{q}}$, executes the following steps:
\begin{enumerate}
\item Each participant $P_{b}^{i_q}$ of $\mathcal{A}_{q}^{s_{i}}$ in $\Gamma_{s_i}$ computes his or her pseudo share  $\mathcal{U}_{P_{b}^{i_q}}=h(x_{P_{b}^{i_q}}||i_u||q_v)$ with the help of his or her share $x_{P_{b}^{i_q}}$ and the publicly available  entities  $h$, $q_v$, $i_u$. 
\item Each member $P_{b}^{i_q}$ sends his or her pseudo share to the combiner.
\end{enumerate} 

\subsection{Combiner Phase}
In this phase the combiner verifies the participants and computes their secret as follows: \\ \\
{\bf(I) Participants Verification Stage}
\begin{enumerate}
\item The combiner receives each of the pseudo share $\mathcal{U}_{P_{b}^{i_q}}$ from each of the participants of $\mathcal{A}_{q}^{s_{i}} \in\Gamma_{s_i}$.
\item The combiner verifies each participant $P_{b}^{i_q}$ by computing $\mathcal{N}_{P_{b}^{iq}}=h(\mathcal{U}_{P_{b}^{i_q}})$, $b=1, 2, \ldots, m_{i_q}$. At this stage, if there be any dishonest participant, he or she will be identified by the combiner. 
\end{enumerate}
{\bf(II) Secret Reconstruction Stage} 
\begin{enumerate}
\item Using the fact that $\mathcal{U}_{P_{b}^{i_q}}+\mathcal{M}_{P_{b}^{i_q}}=\mathcal{B}_{P_{b}^{i_q}}=f_{q}^{s_{i}}(ID_{b}^{i_q})$, the combiner computes the secret $s_i$ as follows \cite{han}\\
$$s_{i}=\sum_{b\in\{1,2,\ldots, m_{i_{q}}\}}(\mathcal{U}_{P_{b}^{i_q}}+\mathcal{M}_{P_{b}^{i_q}})$$ $$\prod_{
\begin{array}{c}
r\in \{1, 2, \ldots,m_{i_{q}}\} 
r \neq b 
\end{array}
}\frac{-ID_{P_{r}^{i_q}}}{ID_{P_{b}^{i_q}}-ID_{P_{r}^{i_q}}} \mod p.$$
\item The combiner sends the secret $s_i$ securely to each of the participants of $\mathcal{A}_{q}^{s_{i}} \in\Gamma_{s_i}$.
\end{enumerate}
\subsection{Participant Phase (II)} 
After getting $s_i$ from the combiner, each participant $P_{b}^{i_q}$  verifies whether the revealed $s_i$ is correct or not with the help of publicly available information $h$ and $S_i$, $ b=1, 2, \ldots, m_{i_{q}}.$ \\

\noindent {\bf Remark:} \\
The verification steps may also be done using the concept of intractability of the discrete logarithm problem over the filed $\mathbb{Z}_p$, instead of using secure collision resistant one way hash function. In that case, for the verification of participants, the dealer may publish $g^{\mathcal{U}_{P_{b}^{i_q}}}$ instead of $h({\mathcal{U}_{P_{b}^{i_q}}})$, where $g$ is a primitive element of the field $\mathbb{Z}_p$. Also for the verification of the combiner, the dealer may publish $g^{s_i}$ instead of $h(s_i)$, $i=1, 2,\ldots, k,$ $ q=1, 2,\ldots , l_{i}$,  $b=1, 2,\ldots, m_{i_{q}}$.  

\section{Renew Process}
\label{1.3}
For more practical use, it is desirable that in an existing secret sharing scheme, secret or secrets, participant or participants and qualified set or sets may be added or may be made inactive by the dealer dynamically without updating the shares of participants of old access structure. In the proposed scheme, that nice property can be incorporated as follows:
\begin{enumerate}
\item {\bf Modification of secret} \\ 
Suppose to the existing scheme, a new secret, say $s_t$, is to be added by the dealer. Let the set of qualified sets of participants corresponding to the secret $s_t$ be $\Gamma_{s_t}=\{\mathcal{A}_1^{s_t}, \mathcal{A}_2^{s_t}, \ldots, \mathcal{A}_{l_t}^{s_t}\}$, where $\mathcal{A}_q^{s_t}= \{P_1^{t_q}, P_2^{t_q}, \ldots, P_{m_{t_q}}^{t_q} \} \subseteq \{ P_1,P_2, \ldots, P_n\}$, $q=1, 2, \ldots, l_{t}$.
The dealer may achieve this as follows:
\begin{itemize}
\item The dealer constructs a new polynomial $f_q^{s_t}$ as described earlier, $q=1, 2, \ldots, l_{t}$.
\item The dealer publishes the values $\mathcal {M}_{P_{b}^{t_q}}$, $\mathcal {N}_{P{_b}^{t_q}}$, and $\mathcal S_t$, where $q=1, 2, \ldots, l_{t}, b=1, 2, \ldots, m_{t_q}$.
\end{itemize}
To make a secret, say $s_t$, inactive corresponding to the access structure $\Gamma_{s_t}$, the dealer may replace that secret by $s_t'$ and may update all the public values related to the secret.

\item {\bf Modification of participant}\\
Suppose, a new participant, say $P_{n+1}$, has to be added by the dealer to the set of participants $\{P_1,P_2, \ldots, P_n \}$. For that, the dealer does the following: 
\begin{itemize}
\item The dealer chooses $x_{n+1} \in_R \mathbb{Z}_{p}$ and $ID_{n+1} \in_R \mathbb{Z}_{p}^*$, distinct from the previously used $x_i$'s and $ID_i$'s for the existing participants. 
\item The dealer sends the value of $x_{n+1}$ securely to the participant $P_{n+1}$. 
\item The dealer publishes the value $ID_{n+1}$ to the public domain.
\item The dealer has to look at the newly formed qualified sets of participants where the new participant $P_{n+1}$ is being added.
\item Modify the pseudo shares as well as the public values of the participants related to the modified qualified sets.
\end{itemize}
If a participant has to be made inactivate by the dealer from the set of participants $\{P_1,P_2, \ldots,$$ P_n \}$, the dealer first has to look at all the access structures, where that participant was present. Then the dealer may change all the secrets of those access structures and update all the public values accordingly.

\item {\bf Modification of qualified set} \\ 
Suppose a new qualified set of participants, say  $\mathcal{A}_{{l_i+1}}^{s_i}$, for the secret $s_i$ is to be added by the dealer to the existing qualified set of participants. To incorporate that, the dealer constructs $f_{l_i+1}^{s_i}$ and publishes the corresponding values $\mathcal {M}_{P_{b}^{i_q}}$, $\mathcal {N}_{P_b^{i_q}}$, where $q=l_{i}+1,~ b=1, 2, \ldots, m_{i_{q}}$. To make  some qualified set of participants inactivate, the dealer has to 
incorporate the similar work as in the case of modification of participants.

\end{enumerate}

\section{Analysis Of The Proposed Scheme}
\label{1.4}

\subsection {Security}
Let us explain the security of the pseudo shares, the shares and the secrets as follows. Here hash function plays an important role.  

\noindent {\bf(I) Security of pseudo Share} \\
An adversary can try to derive participant's pseudo share by using publicly available information $\mathcal{M}_{P_{b}^{i_q}}$. As pseudo share $\mathcal{U}_{P_{b}^{i_q}}$ is protected by $\mathcal{B}_{P_{b}^{i_q}}$, it is not possible for others to derive participant's pseudo share without the knowledge of $\mathcal{B}_{P_{b}^{i_q}}$ which is obtained from the secret polynomial chosen by the dealer. And it is also impossible to adversary to compute $\mathcal{U}_{P_{b}^{i_q}}$ from the previously used pseudo shares (it may happen that set of previously used pseudo shares is empty) of $P_{b}^{i_q}$ under the protection of collision resistant one way hash function. Moreover, it is computationally infeasible to get back $\mathcal{U}_{P_{b}^{i_q}}$ from publicly available entity $\mathcal{N}_{P_{b}^{i_q}}$.

\noindent {\bf(II) Security of Share} \\
Share is chosen by the dealer randomly and delivered to each user $P_j$ secretly. Even though the pseudo share $\mathcal{U}_{P_{b}^{i_q}}$ is compromised, any malicious adversary can not successfully derive $x_j$ from the equation $\mathcal{M}_{P_{b}^{i_q}}=\mathcal{B}_{P_{b}^{i_q}} - \mathcal{U}_{P_{b}^{i_q}}$under the protection of the collision resistant one-way hash function.  \\

\noindent{\bf(III) Security of Secrets} \\
Suppose, all but one participant in $\mathcal{A}_{q}^{s_{i}}$ come to get $s_i$. They have to guess the corresponding value $\mathcal{U}_{P_{b}^{i_q}}$ of the missing participant. As  $\mathcal{U}_{P_{b}^{i_q}}$ is a $[\log_{2}p]+1$ bit long binary string, the forbidden set of participants will have no extra privilege over an outsider who knows only the value of $p$. Moreover, it is computationally infeasible to get back $s_i$ from publicly available entity $S_i$.

From the above discussions, it is clear that the proposed scheme is secure based on the assumption of the hardness of secure collision resistant one-way hash function.

\noindent{\bf Remark:} \\
One of the key parameters concerning the efficiency of a scheme is the size of the share space. In the proposed scheme, the size of the share space is same as that of the secret space.

\subsection {Multi-use}
The proposed scheme is a multi-use one in the sense that the same share of a particular participant may be used to reveal different secrets corresponding to the different qualified sets of participants. This follows form the fact that the pseudo share, submitted to the combiner to reveal a particular secret, of a participant changes for different secrets and even for different qualified subsets of the same secret. This prevents the combiner from misusing the share of a participant to construct some other secrets. 
To incorporate this, successive use of one way hash function was used  in \cite{geng}, \cite{he} and \cite{han}. But in the proposed scheme, successive use of hash function is replaced by the concatenation of suitably chosen binary strings to achieve the same property. This replacement makes the proposed scheme more efficient than above mentioned schemes with respect to the computational cost.

\subsection {Performance Evaluation}
In the proposed scheme, hash function plays the most important role.  So, for a particular secret $s_i$ and one of its corresponding qualified sets $\mathcal {A}^{s_i}_q$, we count the number of times that  the hash function is used by the dealer, the combiner and the members of $\mathcal {A}_q^{s_i}$. In the initialization stage by the dealer, there is no use of hash function. At the time of computing pseudo share, the dealer has to operate hash function $m_{i_q}$ times. In the third stage of the dealer Phase, the dealer has to operate hash function $m_{i_q}+1$ times. So, in the Dealer Phase, hash function is operated $(2m_{i_q}+1)$ times. Each participant operates hash function only once in the Participant Phase (I). When the combiner is going to verify participants,  the combiner has to operate hash function $m_{i_q}$ times. Finally, when participants verify the combiner, then each of them has to operate hash function only once. Note that throughout the protocol, each participant has to operate only hash function just two times and no other operation has required for them. Whereas, in \cite{geng}, \cite{he} and \cite{han} hash function is used successively to reveal more than one secrets when each participant carry only one share. But avoiding the successive use of hash function, we achieve the same goal.

In the Table 1, we highlight the main features of the schemes \cite{geng}, \cite{han}, \cite{he}, \cite{pang}, \cite{wei} along with the proposed scheme in a compact form.

\begin{table}[h]
\begin{center}
\caption{Comparison among \cite{he}, \cite{geng}, \cite{han}, \cite{pang}, \cite{wei} and the proposed scheme with respect to the various parameters. In the following table GAS and PO denote respectively general access structure and predetermined order.}
\label{table1}
\begin{tabular}{|c|c|c|c|}
\hline  & {\it Multi} & {\it Threshold } & {\it Secret } \\
{\it Scheme} &{\it Use} & {\it or } & {\it revealing} \\
& & {\it GAS } & {\it order} \\
\hline He et al. & No & Threshold  & PO\\
\hline Geng et al.& Yes  & Threshold & Any order\\
\hline Han et al. & Yes & Threshold  & Any order\\
\hline Pang et al. & Yes  & GAS  & All secrets \\
& & &   at a time\\
\hline Wei et al. & No  & GAS & PO\\
\hline Proposed & Yes &  GAS  & Any order\\
Scheme & & & \\
\hline
\end{tabular}
\end{center}
\end{table}

\section{\uppercase{Conclusion}}
\label{1.5}
In current scenario, it is important for a secret sharing scheme to be a multi-use, multi-secret, renewable and verifiable for general access structure. The proposed scheme has all the above mentioned properties. Moreover, analysis shows that the proposed scheme is an efficient one and it can provide great capabilities for many applications.

\end{document}